# Guided waves in pre-stressed hyperelastic plates and tubes: Application to the ultrasound elastography of thin-walled soft materials

Guo-Yang Li[1], Qiong He[2], Robert Mangan[3], Guoqiang Xu[1], Chi Mo[1], Jianwen Luo[2], Michel Destrade[3], Yanping Cao[1,*]

[1] *Institute of Biomechanics and Medical Engineering, AML, Department of Engineering Mechanics, Tsinghua University, Beijing 100084, PR China*
[2] *Department of Biomedical Engineering, School of Medicine, Tsinghua University, Beijing 100084, PR China*
[3] *School of Mathematics, Statistics and Applied Mathematics, National University of Ireland Galway, Galway, Ireland*

**Abstract:** *In vivo* measurement of the mechanical properties of thin-walled soft tissues (e.g., mitral valve, artery and bladder) and *in situ* mechanical characterization of thin-walled artificial soft biomaterials in service are of great challenge and difficult to address via commonly used testing methods. Here we investigate the properties of guided waves generated by focused acoustic radiation force in immersed pre-stressed plates and tubes, and show that they can address this challenge. To this end, we carry out both *(i)* a theoretical analysis based on incremental wave motion in finite deformation theory and *(ii)* finite element simulations. Our analysis leads to a novel method based on the ultrasound elastography to image the elastic properties of pre-stressed thin-walled soft tissues and artificial soft materials in a non-destructive and non-invasive manner. To validate the theoretical and numerical solutions and demonstrate the usefulness of the corresponding method in practical measurements, we perform *(iii)* experiments on polyvinyl alcohol cryogel phantoms immersed in water, using the Verasonics V1 System equipped with a L10-5 transducer. Finally, potential clinical applications of the method have been discussed.

---

[*] Corresponding author: Yanping Cao.

Email address: caoyanping@tsinghua.edu.cn; Tel: 86-10-62772520; Fax: 86-10-62781284.

# 1. Introduction

Guided waves in thin-walled structures are widely used in non-destructive testing (NDT) (Achenbach, 2000; Chimenti, 1997; Kim et al., 2006; Raghavan and Cesnik, 2007; Rose, 2002; Su et al., 2006). Understanding the dispersion relations (i.e., the variation of the phase velocities with frequency) of guided waves in the tested material is essential in the NDT and this topic has received considerable attention over the years (Achenbach, 1973; Lowe, 1995; Rose, 2014). Because dispersion relations are sensitive to physical and geometrical parameters, the dispersion features of guided waves can be harnessed for the characterization of materials and structures, e.g., by yielding the elastic moduli, thicknesses and curvatures of elastic shells (Cès et al., 2012; Chimenti, 1997; Moilanen et al., 2007; Yeh and Yang, 2011). To achieve this goal, dispersion curves measured in experiments are fitted with theoretical dispersion curves in order to infer the mechanical and geometrical parameters.

In many circumstances, the tested materials and structures are surrounded by fluid media, e.g., underwater pipelines. It is well known that the dispersion properties of guided waves in fluid-loaded media are also sensitive to the physical properties of the surrounding fluid media. Hence, if the phase velocities of the guided waves are larger than the bulk wave velocities of the surrounding fluid media, the guided waves are the so-called "leaky guided waves" (LGWs), due to energy leakage into the surroundings. Otherwise, the waves are "trapped" within the waveguides (Mazzotti et al., 2014; Rose, 2014). Due to the rich dispersion properties and important engineering applications of fluid-loaded waveguides, the study of guided waves in these structures spans several disciplines (Rose, 2014). For instance, the analytical dispersion relations for waves in a fluid-loaded elastic plate were originally obtained by Osborne and Hart (1945) and have been widely used in the scientific and engineering literature ever since (Aristégui et al., 2001; Chimenti, 1997; Chimenti and Nayfeh, 1985). For waveguides with complicated geometries, numerical methods have been established and validated to calculate the dispersion relations (Hayashi and Inoue, 2014; Mazzotti et al., 2014; Pavlakovic, 1998).

Beyond their use in the characterization of stiff engineering materials, guided waves have also recently been adopted in the ultrasound elastography of *soft thin-walled biological tissues* (Bernal et al., 2011; Couade et al., 2010; Li et al., 2017a; Li et al., 2017b; Nenadic et al., 2016; Nenadic et al., 2011; Urban et al., 2015). The key idea behind ultrasound-based shear wave elastography is to generate elastic waves inside soft biological tissues and then track their propagation with an ultrafast ultrasound imaging method. The elastic properties of the biological soft tissues can then be quantitatively inferred from the measured elastic wave velocities (Bercoff et al., 2004; Jiang et al., 2015; Sarvazyan et al., 1998). In this testing method, the frequencies of the shear waves are usually limited to less than 2 kHz, because of the rapid dissipation of higher frequency shear waves in soft biological tissues (Gennisson et al., 2013; Sarvazyan et al., 2013), and the wavelength is thus of the order of millimeter (recall that bulk shear wave velocities in soft biological tissues are typically 1-10 m/s). For soft biological tissues such as mitral valve, bladder, cornea and artery, the wall thicknesses are smaller than, or comparable to the wavelength, indicating that the elastic waves in these soft tissues are guided by the thin walls. Therefore, guided wave theory (instead of bulk wave theory) should be used in this case to analyze the experimental data and infer their material parameters (Bernal et al., 2011; Couade et al., 2010; Li et al., 2017b).

It is important to note that thin-walled soft tissues and thin-walled artificial soft biomaterials in their working state are usually subject to *pre-stress*. Although the effects of pre-stresses on the dispersion relations of guided waves in elastic materials have been systematically investigated in the literature (Ogden and Roxburgh, 1993; Rogerson and Fu, 1995; Bagno and Guz', 1997; Kaplunov et al., 2000, 2002; Wijeyewickrema and Leungvichcharoen, 2009; Kayestha et al., 2010; Akbarov et al., 2011), guided waves in *fluid-loaded* pre-stressed thin-walled soft materials have been less studied (see for example, Bagno and Guz', 1997; and references therein). Compared with hard engineering materials, soft materials can undergo large deformations in response to external loads, which influences greatly the characteristics of the elastic waves (Destrade and Ogden, 2010; Gennisson et al., 2007; Jiang et al., 2015). In this sense, to address the effects of pre-stresses on the dispersion properties of guided waves in thin-walled soft materials, it is necessary and important to conduct the analysis within the framework of finite deformation theory.

Based on the premise above, this paper investigates theoretically, numerically, and experimentally, the propagation of guided waves in pre-stressed soft plates and tubes surrounded by fluid. The results may serve as fundamental solutions to characterize the mechanical properties of thin-walled soft materials and soft tissues including mitral valve, artery, cornea and bladder, using the ultrasound elastography method.

The paper is organized as follows. A theoretical analysis is conducted in Section 2 based on the incremental theory of nonlinear elasticity (Ogden, 1984; Ogden, 2007). The analytical dispersion relations for both the anti-symmetric and symmetric modes of the guided waves in a pre-stressed hyperelastic plate immersed in inviscid fluid are derived. In Section 3, a finite element model is built to validate the analytical solutions and demonstrate their applicability in describing the dispersion relations of the guided circumferential waves in a tube. In Section 4, phantom experiments are carried out on pre-stretched soft plates immersed in fluid to verify the theoretical solutions and to demonstrate the usefulness of the corresponding method in practical measurements. Section 5 discusses the potential applications of the method in medical image analysis. Section 6 concludes the paper.

## 2. Theoretical analysis

We consider that thin-walled soft tissues in their *in vivo* state, and thin-walled artificial soft biomaterials in their working state, are all subject to pre-stress. In this section, we first derive analytically the dispersion relations of guided waves in pre-stressed hyperelastic plates immersed in an inviscid fluid (Fig. 1). Then we extend our analysis to curved plates to include tubular geometries. To address the effect of pre-stresses on the dispersion relations, we rely on the incremental theory of nonlinear elasticity (Ogden, 1984; Ogden, 2007). In the following section, we recall the basic equations for wave motion in pre-stressed hyperelastic plates and inviscid fluids.

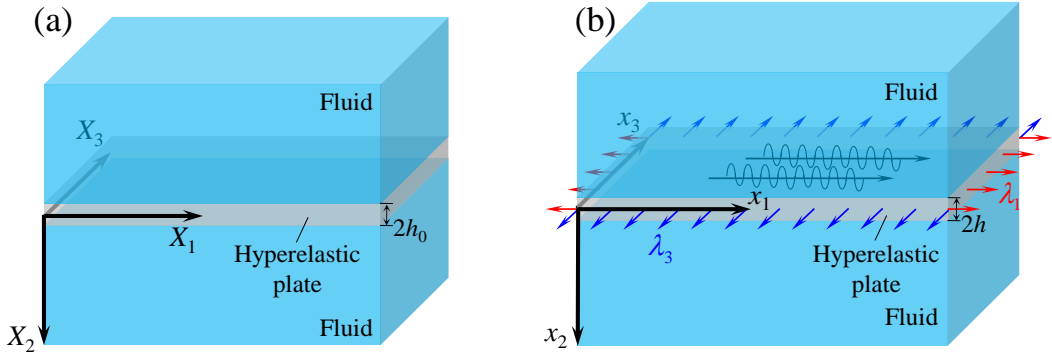

**Fig. 1** Sketch of the model. The elastic plate immersed in fluid is pre-stretched homogeneously, which causes a finite deformation of the plate from **(a)** the reference configuration to **(b)** the current (or deformed) configuration. Then, small amplitude elastic wave propagation is studied in the current configuration.

## 2.1. Governing equations for waves in the pre-stressed hyperelastic plate

### 2.1.1. The plate

The soft plate is modelled as an incompressible isotropic hyperelastic solid, characterized by its mass density $\rho$ and strain energy $W$. It is subject to a pre-stress $\boldsymbol{\sigma}$ and maintained in a state of homogeneous deformation characterized by the principal stretch ratios $\lambda_1$, $\lambda_2$, $\lambda_3$, along the principal Eulerian axes $x_1$, $x_2$, $x_3$. Then $W = W(\lambda_1, \lambda_2, \lambda_3)$, subject to $\lambda_1 \lambda_2 \lambda_3 = 1$. The dimension of the plate in the $x_1, x_3$ plane is much greater than its thickness. Initially, it is of thickness $2h_0$, and in its deformed state, of thickness $2h = 2\lambda_2 h_0$. This assumption on the deformation state is acceptable in the region of interest (ROI) with finite dimension in ultrasound elastography (Jiang et al., 2015).

Then we can use the equations of incremental elasticity to describe the propagation of a small-amplitude wave in $(x_1, x_2)$ plane with displacement $\boldsymbol{u} = \boldsymbol{u}(x_1, x_2, t)$. Because of incompressibility, we have $u_{1,1} + u_{2,2} = 0$, so that we may introduce a stream function $\psi$ such that

$$u_1 = \psi_{,2}, \quad u_2 = -\psi_{,1}, \tag{1}$$

It can be shown that the propagation of the plane wave in the pre-stressed solid is governed by the following equation of motion (Ogden, 2007)

$$\alpha \psi_{,1111} + 2\beta \psi_{,1122} + \gamma \psi_{,2222} = \rho(\psi_{,11tt} + \psi_{,22tt}), \tag{2}$$

and that the incremental nominal traction components are $\dot{S}_{021}$ and $\dot{S}_{022}$, found from

$$\dot{S}_{021} = (\sigma_{22} - \gamma)\psi_{,11} + \gamma \psi_{,22}, \tag{3}$$

$$\dot{S}_{022,1} = \rho\psi_{,2tt} - (2\beta + \gamma - \sigma_{22})\psi_{,112} - \gamma\psi_{,222}. \tag{4}$$

Here the acousto-elastic coefficients $\alpha, \beta, \gamma$ are given by

$$\alpha = \frac{\lambda_1^2\left(\lambda_1\frac{\partial W}{\partial \lambda_1} - \lambda_2\frac{\partial W}{\partial \lambda_2}\right)}{\lambda_1^2 - \lambda_2^2}, \quad \gamma = \frac{\lambda_2^2}{\lambda_1^2}\alpha, \tag{5}$$

$$2\beta = \lambda_1^2\frac{\partial^2 W}{\partial \lambda_1^2} - 2\lambda_1\lambda_2\frac{\partial^2 W}{\partial \lambda_1 \partial \lambda_2} + \lambda_2^2\frac{\partial^2 W}{\partial \lambda_2^2} - 2\frac{\lambda_1\lambda_2\left(\lambda_2\frac{\partial W}{\partial \lambda_1} - \lambda_1\frac{\partial W}{\partial \lambda_2}\right)}{\lambda_1^2 - \lambda_2^2}.$$

To model the hyperelastic deformation behavior of the plate we will use in turn the *Fung-Demiray* strain energy density (Demiray, 1972; Fung et al., 1979) for isotropic soft tissues,

$$W = \frac{\mu_0}{2b}\left(e^{b(I_1-3)} - 1\right), \tag{6}$$

where $I_1 = \lambda_1^2 + \lambda_2^2 + \lambda_3^2$, $\mu_0 > 0$ is the initial shear modulus and $b > 0$ is a hardening parameter (Demiray, 1972); the *neo-Hookean* model,

$$W = \frac{\mu_0}{2}(I_1 - 3), \tag{7}$$

(which corresponds to $b=0$ in the Fung-Demiray model), and the general fourth-order model of weakly non-linear elasticity

$$W = \mu_0 i_2 + \frac{A}{3}i_3 + D(i_2)^2, \tag{8}$$

where $i_k = [\sum_{i=1}^{3}(\lambda_i^2 - 1)^k]/2^k$, $k = 2,3$, and $A, D$ are the Landau coefficients of non-linearity (Destrade and Ogden, 2010).

For those three models, the acousto-elastic coefficients are easily computed from the formulas (5). For instance, in the case of the neo-Hookean solid (7), they are

$$\alpha = \mu_0\lambda_1^2, \quad \beta = \frac{\mu_0}{2}(\lambda_1^2 + \lambda_2^2), \quad \gamma = \mu_0\lambda_2^2. \tag{9}$$

### 2.1.2. *The fluid*

The surrounding fluid is modelled as compressible and inviscid, and we assume that the motion is irrotational. In the static state, the fluid is subject only to a hydrostatic pressure $\boldsymbol{\sigma}^* = -P\boldsymbol{I}$. However, because continuity of stress is required across the fluid-solid boundaries in the reference configuration, we must have $P = -\sigma_{22}$.

The governing equation of motion for the small displacement of the fluid reads

$$\nabla(\kappa\nabla \cdot \boldsymbol{u}^F) = \rho^F\ddot{\boldsymbol{u}}^F, \tag{10}$$

where $\kappa$ and $\rho^F$ denote the bulk modulus and mass density of the fluid, respectively, and $\boldsymbol{u}^F$ is the mechanical displacement of the fluid, measured in the current configuration. Because the motion is irrotational, we have $\nabla \times \boldsymbol{u}^F = 0$, and hence the displacement can be written in the form $\boldsymbol{u}^F = \nabla\chi$, where $\chi = \chi(x_1, x_2, t)$ is a scalar potential. Using this expression, Eq.(10) can be rewritten as

$$\nabla^2 \chi = \frac{1}{c_p^2} \ddot{\chi} \tag{11}$$

where $c_p = \sqrt{\kappa/\rho^F}$ is the speed of sound in the fluid. The incremental pressure in the fluid induced by the deformation is

$$p = -\kappa \nabla \cdot \boldsymbol{u}^F. \tag{12}$$

### 2.1.3. The interface

Let $\boldsymbol{n}$ denote the unit outward normal of the plate. The interfacial conditions between the pre-stressed plate and the fluid are (Otténio et al., 2007)

$$\dot{\boldsymbol{S}}_0^T \boldsymbol{n} = -p\boldsymbol{n} - \sigma_{22}(\nabla \boldsymbol{u})^T \boldsymbol{n}, \tag{13}$$

and

$$\boldsymbol{u} \cdot \boldsymbol{n} = \boldsymbol{u}^F \cdot \boldsymbol{n}, \tag{14}$$

from which we find the following boundary conditions in component form

$$u_2 = u_2^F, \quad \dot{S}_{021} = -\sigma_{22} u_{2,1}, \quad \dot{S}_{022,1} = -p_{,1} - \sigma_{22} u_{2,21}, \tag{15}$$

at $x_2 = \pm h$.

### 2.2. Dispersion analysis of the guided waves

We now consider a guided wave, travelling in the $x_1$-direction with angular frequency $\omega$, wave-number $k$ and speed $c = \omega/k$, which is attenuated with distance away from the plate. Hence we seek a wave solution of the form

$$\chi = A e^{rkx_2} e^{ik(x_1 - ct)}, \quad \psi = B e^{skx_2} e^{ik(x_1 - ct)}, \tag{16}$$

where $r$ and $s$ are attenuation factors, and $A, B$ are constants.

In the fluid we find from Eq.(11) that

$$r^2 - 1 = -\frac{c^2}{c_p^2}. \tag{17}$$

The solutions to Eq. (17) are $r = \pm \xi$, where $\xi = \sqrt{1 - \frac{c^2}{c_p^2}}$. It follows that the solution for the fluid in the region $x_2 \geq h$ ($x_2 \leq -h$, respectively) is

$$\chi^{\pm} = A^{\pm} e^{(\mp \xi k x_2)} e^{ik(x_1 - ct)}. \tag{18}$$

In the plate we find from Eq.(2) that

$$\gamma s^4 - (2\beta - \rho c^2) s^2 + \alpha - \rho c^2 = 0. \tag{19}$$

We call $s_1^2, s_2^2$ the roots of this quadratic in $s^2$, and conclude that

$$\psi = \phi(x_2) e^{ik(x_1 - ct)}, \tag{20}$$

with

$$\phi(x_2) = B_1 \cosh(s_1 k x_2) + B_2 \sinh(s_1 k x_2) + B_3 \cosh(s_2 k x_2) + B_4 \sinh(s_2 k x_2), \tag{21}$$

where $B_1, B_2, B_3, B_4$ are constants.

Similar to the analysis for Lamb waves in plates surrounded by vacuum (Ogden and Roxburgh, 1993), we find that anti-symmetric modes ($B_2 = B_4 = 0$) decouple from symmetric modes ($B_1 = B_3 = 0$). In contrast to that analysis, we find that $\sigma_{22}$ disappears from the equations once the boundary conditions are applied, a hallmark of waves at the interface between two media (Otténio et al., 2007). For instance, for the anti-symmetric mode, the boundary conditions (15) written at $x_2 = \pm h$ first show that $A^+ = A^- = A$ (say) and then read

$$\cosh(s_1 kh) B_1 + \cosh(s_2 kh) B_3 + i\xi e^{-\xi kh} A = 0, \tag{22}$$
$$(1 + s_1^2)\cosh(s_1 kh) B_1 + (1 + s_2^2)\cosh(s_2 kh) B_3 = 0, \tag{23}$$
$$\gamma s_1 (1+s_2^2) \sinh(s_1 kh) B_1 + \gamma s_2 (1+s_1^2) \sinh(s_2 kh) B_3 + i\rho^F c^2 e^{-\xi kh} A = 0, \tag{24}$$

where we used the identity $2\beta - \rho c^2 = \gamma(s_1^2 + s_2^2)$.

This homogeneous system has non-trivial solutions for the amplitudes $B_1, B_3, A$ when its determinant is zero, which is the *dispersion equation*:

$$\gamma s_1 (1+s_2^2)^2 \tanh(s_1 kh) - \gamma s_2 (1+s_1^2)^2 \tanh(s_2 kh) + \frac{\rho^F c^2}{\xi}(s_1^2 - s_2^2) = 0. \tag{25}$$

When there is no fluid around the plate, $\rho^F = 0$ and we recover the dispersion equation of (Ogden and Roxburgh, 1993) for a pre-deformed plate in vacuum (with $\sigma_{22} = 0$ in their equation). When the plate is not stretched $\alpha = \beta = \gamma = \mu_0$, so that $s_1^2 = 1, s_2^2 = 1 - \frac{\rho c^2}{\mu_0}$ and the dispersion equation simplifies to

$$\left(2 - \frac{\rho c^2}{\mu_0}\right)^2 \tanh(kh_0) - 4\sqrt{1 - \frac{\rho c^2}{\mu_0}} \tanh\left(\sqrt{1 - \frac{\rho c^2}{\mu_0}} kh_0\right) + \frac{\rho \rho^F c^4}{\mu_0^2 \sqrt{1 - \frac{c^2}{c_p^2}}} = 0 \tag{26}$$

in agreement with the linear elasticity result of (Osborne and Hart, 1945).

Finally, for symmetric modes, the roles of Cosh and Sinh are interchanged in the boundary conditions, and the dispersion equation is (25) where Tanh is replaced with Cotanh.

To plot the dispersion curves, we fix the circular frequency $f = \omega/2\pi$ (noticing $k = \omega/c$) and solve numerically the dispersion equation for the speed $c$. There are infinitely many solutions to the dispersion equation, which correspond to the different branches of the anti-symmetric and symmetric modes (denoted by $A_n, S_n, n = 0,1,2,...$). For illustration, we solve the dispersion equation for the neo-Hookean plate, and show the numerical results in Fig. 2: solid lines show the results without pre-stress ($\lambda_1 = \lambda_2 = \lambda_3 = 1$) and dashed lines show the dispersion curves with a homogeneous pre-stress leading to $\lambda_1 = 1.1, \lambda_2 = 1/1.1, \lambda_3 = 1$. Curves in different colors represent different modes, but in the present study we concentrate on the lowest anti-symmetric mode $A_0$ and symmetric mode $S_0$ (the fundamental modes) which are predominant in the low frequency range.

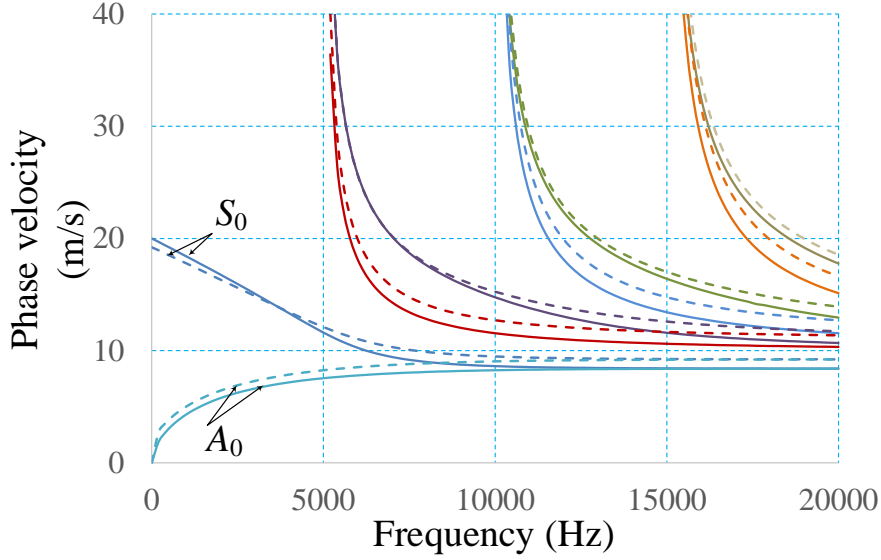

**Fig. 2** Dispersion curves for the neo-Hookean plate without pre-stress (solid lines) and with pre-stress (dashed lines) when $\lambda_1 = 1.1, \lambda_2 = 1/1.1, \lambda_3 = 1$. The initial shear modulus, mass density and thickness of the plate are 100 kPa, 1,000 kg/m³ and 1 mm, respectively.

Our analytical solution reveals the effect of the state of pre-deformation on the dispersion relation. The state of homogeneous deformation of the plate is entirely determined by the principal stretch ratios $\lambda_i$ ($i = 1,2,3$). Defining $\lambda_1 = \lambda$, and $\lambda_2 = \lambda^{-\zeta}$, then we obtain $\lambda_3 = \lambda^{\zeta-1}$ from the constraint $\lambda_1 \lambda_2 \lambda_3 = 1$ for incompressible materials. The parameter $\zeta$ is in the range of 0.5 to 1 and determined by the deformation state of the plate: for example $\zeta = 1$ describes plane strain and $\zeta = 0.5$ describes uni-axial tension. For the neo-Hookean model (7) and Fung-Demiray model (6) with $b = 5$, we study the variation of the $A_0$ and $S_0$ modes with $\zeta$. From Fig. 3, it is interesting to find that $\zeta$ basically has no effect on the $A_0$ mode, as the curves are undistinguishable one from another especially when the parameter $b$ is small (for neo-Hookean model, $b = 0$). This can be explained as that for a smaller $b$ (the material hardening effect is less significant), when the stretch ratio along the wave propagation direction is prescribed, varying the parameter $\zeta$ will not significantly change the material stiffness. The dispersion relation of the $A_0$ mode is insensitive to $\zeta$ indicates that only the principal stretch ratio $\lambda$ along the wave propagation direction needs to be measured to predict the effect of pre-stress on the $A_0$ mode using our analytical solution. This finding brings great ease for the practical application of our analytical solution because the $A_0$ mode is used in the ultrasound elastography of thin-walled soft biological tissues (Bernal et al., 2011; Li et al., 2017a) and it is not easy to accurately evaluate the parameter $\zeta$ in practical measurements.

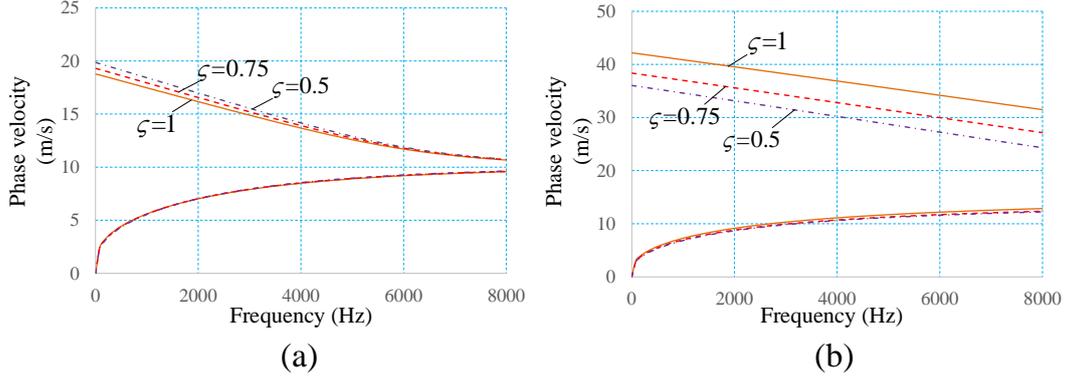

**Fig. 3** Effect of the parameter $\zeta$ on the $A_0$ and $S_0$ modes for plates with a 20% extension ($\lambda = 1.2$). Here $\zeta = 1$ (solid line), $\zeta = 0.75$ (dashed line), $\zeta = 0.5$ (dash-dot line) for: **(a)** neo-Hookean model; **(b)** Fung-Demiray model with $b = 5$. The initial shear modulus and wall thickness of the plate are 100 kPa and 1 mm, respectively.

## 3. Finite element simulations

To validate the analytical solutions obtained in Section 2 and investigate the guided waves in a curved plate or tubular structures, we establish Finite Element (FE) models to calculate the dispersion relations from numerical simulations.

### 3.1. Validation of the theoretical solutions

A plane strain model as shown in Fig. 4(a) is firstly established to validate the analytical solutions. The model consists of three parts: fluid/plate/fluid. The interfacial conditions between the fluid and the plate given by Eqs. (13) and (14) are realized with the 'tie' constraint in ABAQUS (2010).

In the simulation, the model is firstly stretched from its initial length $L_0$ with $L_0 = 30h_0$ to $L = \lambda L_0$ in order to pre-stress the plate, and during this process, the arbitrary Lagrangian-Eulerian (ALE) mesh technique is adopted to avoid mesh distortion in the fluid region, which is modeled as the acoustic medium (AC2D8) (ABAQUS, 2010). Hybrid elements (CPE8RH) are adopted to model the incompressible plate and the number of the elements along thickness direction is 12.

Then the dispersion curve is calculated on the deformed configuration with the frequency domain FE method. In this step, the following periodic boundary conditions (PBCs) are invoked in the simulations

$$\mathbf{u}^{BC} = \mathbf{u}^{B'C'}, \quad p^{AB} = p^{A'B'}, \quad p^{CD} = p^{C'D'}, \tag{27}$$

where the superscript indicates the boundary region as shown in Fig. 4(b). Those constrain equations can be realized by multi-point constrains (MPCs). The natural frequency analysis is conducted. Supposing that for a specified modal shape, the frequency is $f_n$, where $n = 1,2,3,...$ is the number of waves between two sides of the model. The corresponding wavelength is $l_n = L/n$ and the phase velocity can be determined as

$$c_n = l_n f_n. \tag{28}$$

For illustration, Figs. 4(c)-(d) show the modal shape of the plate and the corresponding pressure field in the fluid for the anti-symmetric mode and the symmetric mode when $n=3$. Taking the stretch ratio $\lambda$ as 1.0 (no pre-stress) and then 1.1 (10% extension), the dispersion curves for the fundamental anti-symmetric and symmetric modes are plotted in Fig. 5(a) for the neo-Hookean model (7) and in Fig. 5(b) for the Fung-Demiray model (6) with $b=1$. The solid and the dashed lines are the theoretical solutions and the discrete points are the FE results, and they match each other very well. Clearly, the pre-stretch increases the phase velocities of both the anti-symmetric and symmetric modes, and the higher the frequency, the more significant the increase. Similarly, the greater the hardening parameter $b$ is (going from $b=0$ to $b=1$), the larger the increase in phase velocities will be, especially for the symmetric mode. When the frequency tends to infinity, the phase velocities of the anti-symmetric and the symmetric modes both tend to the same phase velocities of the (half-spaces) interfacial waves, consistent with previous studies (Otténio et al., 2007).

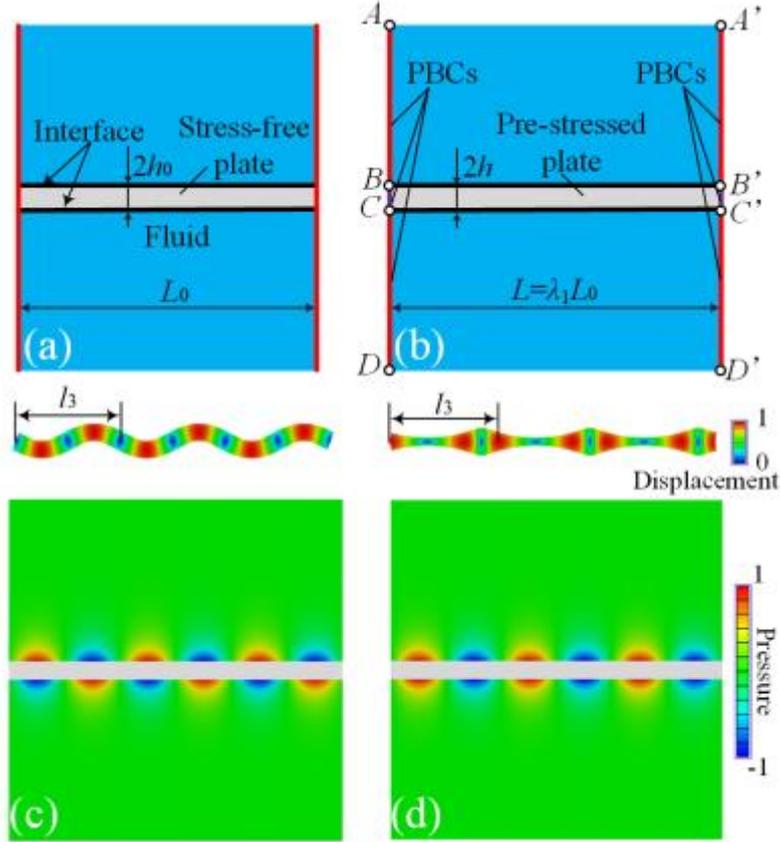

**Fig. 4** The FE model used to calculate the dispersion relations. **(a)-(b)** The plate is pre-stretched. The dispersion relation is calculated based on the deformed configuration by using the PBCs. The resulting modal shape of the plate and corresponding pressure field in the fluid for: **(c)** anti-symmetric mode; and **(d)** symmetric mode. Here the number of the waves between the two sides of the model is $n=3$.

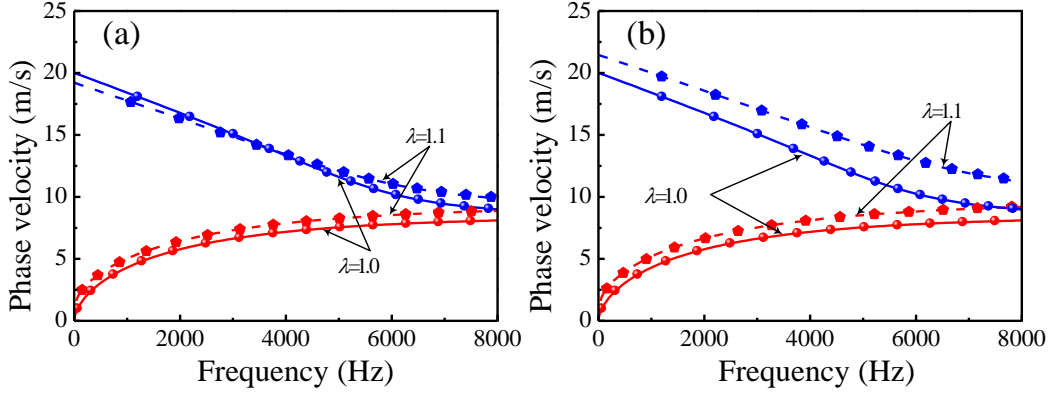

**Fig. 5** Dispersion curves for the fundamental anti-symmetric (lower curves) and symmetric (upper curves) modes: Comparison of the theoretical solutions (solid and dashed lines) with FE results (discrete points) when **(a)** $b = 0$ (neo-Hookean model); **(b)** $b = 1$ (Fung-Demiray model), and when the plate is unstretched ($\lambda = 1.0$) and stretched by 10% ($\lambda = 1.1$). The initial thickness, the initial shear modulus and the mass density of the plate are $2h_0 = 1$mm, $\mu_0 = 100$ kPa, and $\rho = 1,000$ kg/m³, respectively, and the bulk modulus and mass density of the fluid are $\kappa = 2.2$ GPa and $\rho^F = 1,000$ kg/m³, respectively.

### 3.2. Guided circumferential waves (GCWs) in pre-stressed tubes

For guided waves in curved structures, such as circular cylindrical tubes, explicit dispersion equations cannot be obtained and a numerical integration of the equations of motion through the wall thickness is required, for instance based on the Stroh formulation and the surface impedance method (Shuvalov, 2003). Here we do not need that treatment, because it is found from our numerical simulations, that when the ratio between the wall thickness and the radius of curvature is smaller than a critical ratio, the dispersion equation of the curved structure can be accurately approximated by that of flat plate established in the previous section, similar to the result for the stress-free tube reported by Fong (2005) and our previous study (Li et al., 2017b).

We consider that the tube is pre-stressed by the internal pressure, which is $P$ mmHg higher than that of the external pressure. As shown in Fig. 6(b), the inner radius and wall thickness of the tube are $R_0$ and $2h_0$ in the stress-free configuration, and $R$ and $2h$ in the deformed configuration, respectively. The dispersion graphs for this case are found with the FE method and results are shown in Fig. 7 (points). We see that the phase velocities significantly increase with an increase in the pressure $P$.

To compare the dispersion curves obtained from the FE analysis for tubes to those obtained with a plane geometry, we consider a plate as shown in Fig. 6 (c), which is totally constrained along $x_3$-direction and pre-stretched along the $x_1$-direction with the prescribed stretch ratio $\lambda_{\theta\theta}$. Here $\lambda_{\theta\theta}$ is the average circumferential stretch ratio in the tube shown in Fig. 6(b), determined by

$$\lambda_{\theta\theta} = \frac{R + h}{R_0 + h_0}. \tag{29}$$

Different pressure levels $P$ lead to different curvature radii $R$ and wall thicknesses $2h$, and the corresponding stretch ratio $\lambda_{\theta\theta}$ can be calculated from Eq.(29). Then the anti-symmetric mode of the dispersion curve for an immersed plate can be calculated with Eq.(25).

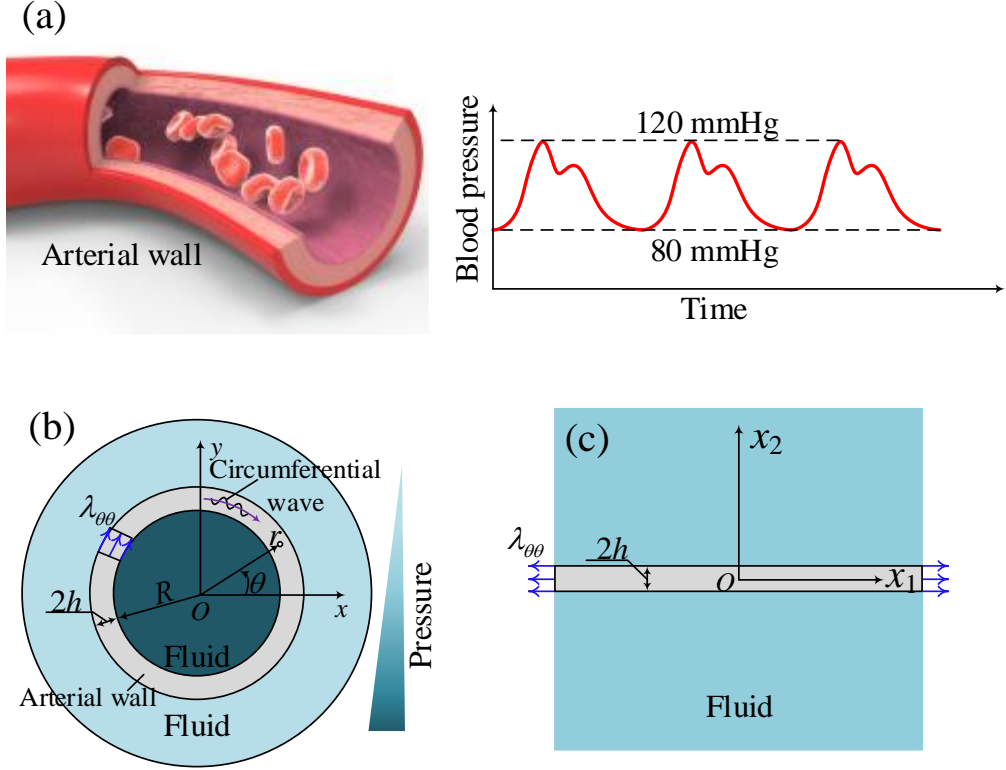

**Fig. 6** **(a)** The arterial wall (image credit: Boston Children's Hospital's Science and Clinical innovation blog, website: https://vector.childrenshospital.org/) and the variation of the blood pressure; **(b)** Guided circumferential wave in a pre-stressed tube subjected to internal fluid pressure $P$; and **(c)** the corresponding approximation model with plane geometry, when the wall thickness is small compared to the radius of curvature.

The theoretical predictions of the dispersion curves are shown in Fig. 7 (lines). We see that the theoretical solution given by Eq.(25) derived for the pre-stressed flat plate with the stretch ratio given by Eq.(29) can describe well the dispersion properties of the GCWs in a specified frequency range, e.g., 100 Hz to 3000 Hz. The curvature effects of the tube are dictated by the ratio $R_0/(2h_0)$, and the larger it is, the less it affects the dispersion curves. In the present example, $R_0/(2h_0) = 2$, which is a common value for some arterial walls, e.g., carotid artery (for the bladder it is a much larger number). The results in Fig. 7 indicate that at least for $R_0/(2h_0) \geq 2$, the dispersion curves of the GCWs in a pre-stressed tube can be very well approximated with those of a flat plate. This finding indicates that our analytical treatment of Section 2 can be readily used in the ultrasound elastography of arteries. For example, consider an *in vivo* arterial wall as shown in Fig. 6(a), subject to a variable blood pressure (typically 80-120 mmHg). Our present analytical solution enables quantitative understanding of the effect of blood pressure on the dispersion curves which are used in elastography of the arteries in order to infer the elastic properties of arterial wall.

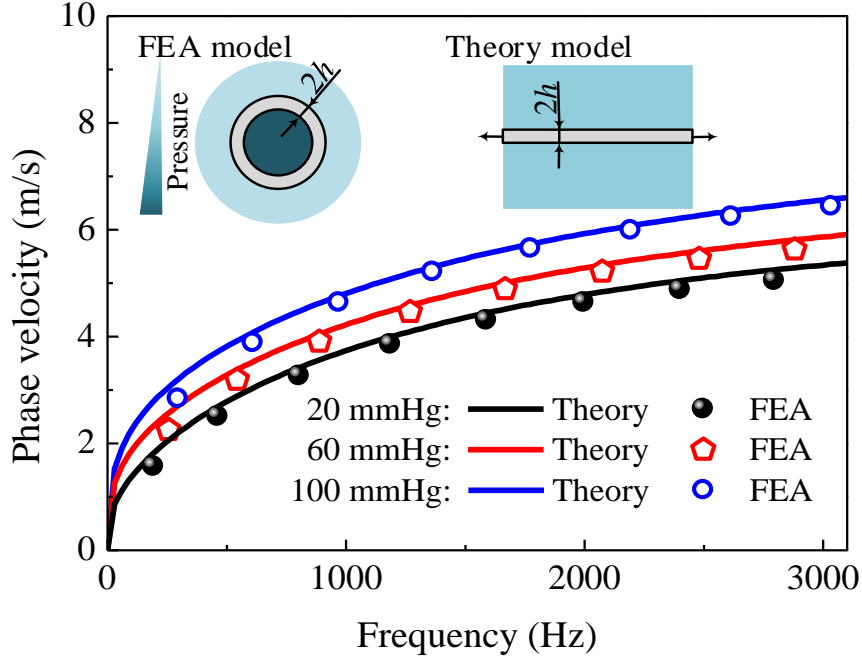

**Fig. 7** Dispersion curves (discrete points) of guided circumferential waves in a pre-stressed tube made of Fung- Demiray material, as found from FE analysis and dispersion curves (solid lines) of the approximation model given by Eq.(25). The pressure $P$ varies from 20 to 100 mmHg and the hardening parameter is $b = 1.0$.

## 4. Experiments on phantom gels

Our above theoretical and numerical analyses lead the way to an ultrasound-based *elastography method* to infer the mechanical properties of pre-stressed thin-walled soft tissues and artificial thin-walled soft biomaterials in their working state. To validate the method and demonstrate its usefulness in practical measurements, we performed experiments on phantom gels, as reported in this section.

We prepared a polyvinyl alcohol (PVA) cryogel phantom as follows. The PVA solution was made of 10% (by weight) PVA (Sigma-Aldrich, Shanghai, China), 87% distilled water, and 3% Sigmacell cellulose (20 μm, Sigma-Aldrich, Shanghai, China). The latter provided ultrasound scattering particles. The mixture was kept at a temperature of 85°C and stirred until the powder was fully dissolved. Then the mixture was cooled and underwent four freeze/thaw (F/T) cycles with 12h of freezing (-20 °C) and 12h of thawing (20 °C). After that, the phantom was mounted on a tensile machine to introduce pre-stress by prescribing the stretch ratio. The local deformation in the phantom along the stretch direction, as quantified by $\lambda$, was determined from the displacements of the grids marked on the surface of the phantom as shown in Fig. 8(a). According to the finding in Fig. 4, we may simply take $\zeta = 1$ in our data analysis, because the actual value of $\zeta$ has negligible effect on the $A_0$ mode used in our method.

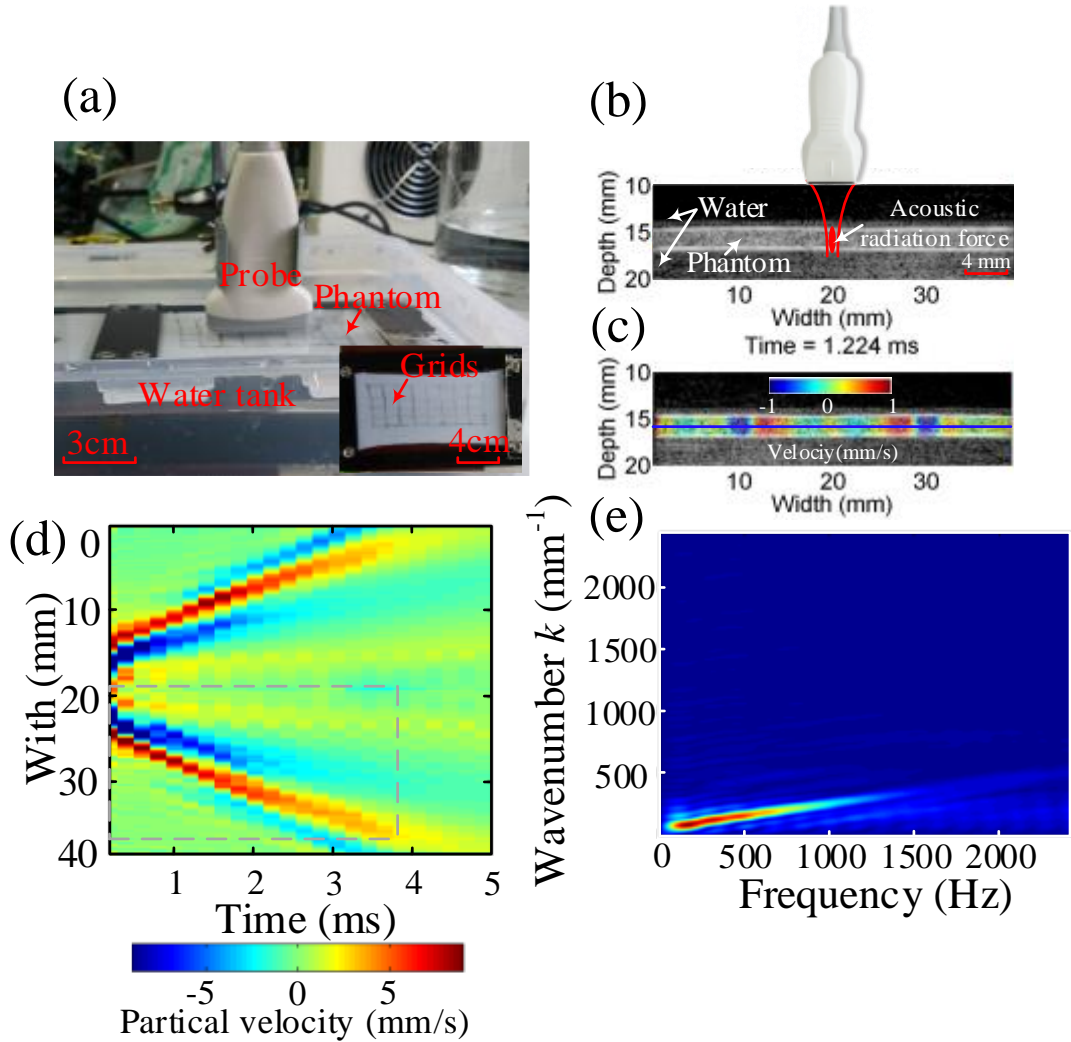

**Fig. 8** (a) Setup of the experiments on phantom gel. The phantom was stretched with a tensile machine and immersed in water. Grids were marked on its surface to evaluate the strains. (b) A typical B-mode image of the gel. (c) The velocity field at 1.224 ms after push. The wave propagated from the center of the phantom in two opposite directions. (d) Along the blue line (see (c)), the spatio-temporal imaging of the waves is shown. (e) The magnitude map of the results obtained by conducting 2D-Fourier Transforms to (d). At each frequency, the corresponding wavenumber is identified by finding the maximum value of the magnitude.

In our experiments, the phantom was immersed in water. The Verasonics V1 System (Verasonics Inc., Kirkland, WA, USA) equipped with a L10-5 transducer (Jiarui, Shenzhen, Guangdong, China, $f_0$ =7.5 MHz,) was used to generate elastic waves in the phantom, i.e., push the PVA phantom by applying the acoustic radiation force produced by the momentum transfer from the acoustic waves to the PVA phantom (Torr, 1984), and acquire the in-phase and quadrature (IQ) data. A typical B-mode image and the velocity field at 1.224 ms after push are shown in Fig. 8(b) and (c), respectively. Fig. 8(d) gives the spatio-temporal imaging of the waves along the blue line shown in Fig. 8(c). Accordingly, the dispersion curves can be obtained by conducting the two-dimensional Fourier Transformation (2DFT) to Fig. 8(d) (Alleyne and Cawley, 1991; Li et al., 2017b). Briefly, from the magnitude map obtained by conducting 2DFT to the spatio-temporal image, i.e., Fig. 8(e), the wavenumber can be identified by finding the maximum value of the magnitude at each frequency. We then fitted the experimental dispersion curves with the theoretical dispersion relation Eq.(25), using the neo-Hookean (7), Fung-Demiray (6), and fourth-order elasticity (8) models to describe the PVA phantom.

The parameter optimization was done using the curve fitting function *curve_fit* and the root finding function *fsolve* in the Python module SciPy. The curve fitting function *curve_fit* only accepts an explicitly defined function as its argument. Since Eq.(25) defines the speed $c$ only implicitly, it was necessary to define a Python function which, given the angular frequency $\omega$ and the material parameters of the chosen hyperelastic model, returns the speed $c$ as the output. This function was established by finding numerically the root of the dispersion equation Eq.(25).

In the experiments, three different stretch ratios were prescribed. In the absence of pre-stretch, i.e. $\lambda = 1.0$, the initial shear modulus $\mu_0$ was first inferred by fitting the experimental dispersion curve with Eq.(26). As shown in Fig. 9 (a), the initial shear modulus was found to be $\mu_0 = 37$ kPa. This value is consistent with the previous results (Li et al., 2017a; Li et al., 2017b) and we further validated it by conducting a separate tensile test, which yielded $\mu_0 = 36$ kPa, see Fig. 9(b).

The shear modulus $\mu_0$ is the only material parameter of the *neo-Hookean model*; once it has been determined, the dispersion equation Eq.(25) can be further used to predict the dispersion curves for different prescribed stretch ratios. Fig. 9(a) clearly shows that the theoretical predictions match the experimental dispersion curves well when the phantom has been stretched by 8% and then by 18%. This indicates that the hardening effects of the phantom material are not very significant and for this type of materials the present method can be applied to infer the material parameter of the neo-Hookean model of pre-stressed thin-walled soft materials once the value of the pre-stress is provided or estimated.

For the *Fung-Demiray model*, we fixed $\mu_0 = 37$ kPa as determined from the $\lambda = 1.0$ curve and then used the curve with the highest stretch $\lambda = 1.18$ to determine the harnding parameter $b$ from the optimization procedure, and found $b = 0.22$, indicating that indeed the hardening effect of the material is not significant. Fig. 9(c) shows that the corresponding model provided a good predictive fit for the intermediate curve at $\lambda = 1.08$.

A similar approach can be used for the *fourth-order elasticity model* to determine $A$ and $D$. Here, however, we took the point of view that the dispersion curve when the material was unstretched was unknown, as would be the case *in situ*. Hence we determined all three material parameters ($\mu, A$ and $D$) by fitting to the curve of $\lambda = 1.18$ curve. We found $\mu_0 = 38$ kPa, $A = -126$ kPa, $D = 23$ kPa. These values provided very good predictive fits for both the $\lambda = 1.00$ and $\lambda = 1.08$ curves, see Fig. 9(d).

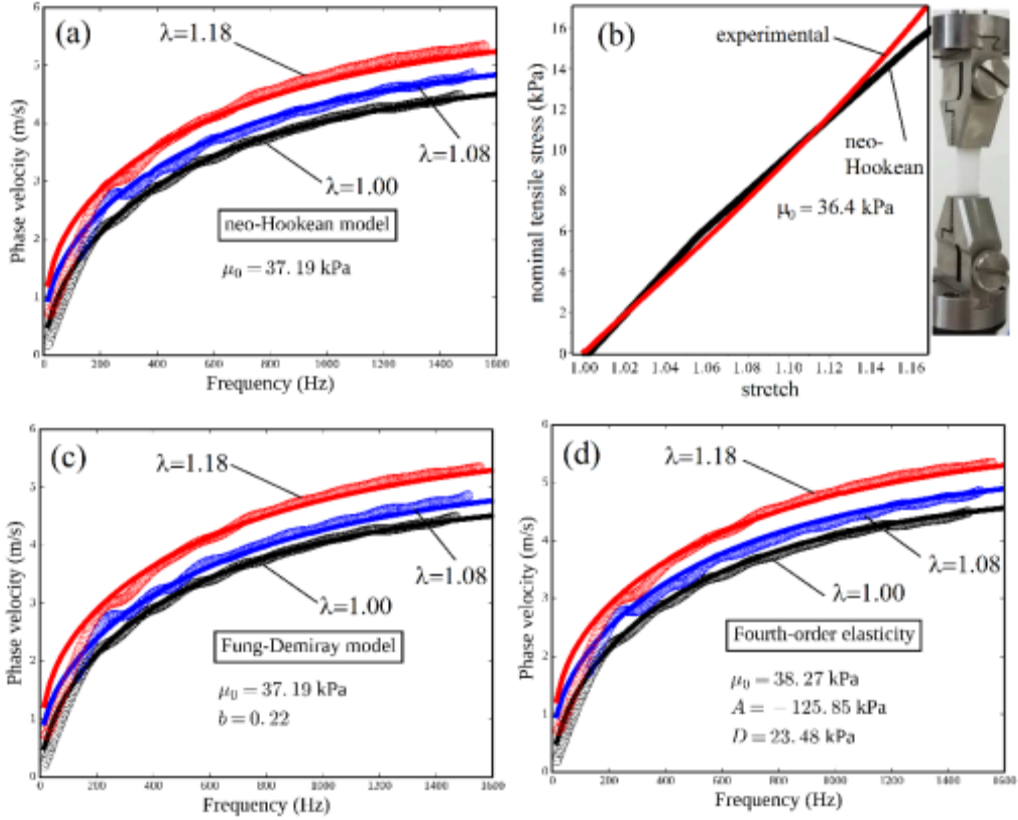

**Fig. 9** Experimental (points) and theoretical (lines) dispersion curves at different stretch ratios for the PVA gel. The initial thickness and mass density of the plate are $2h_0 = 2$ mm and $\rho = 1.1 \times 10^3$ kg/m$^3$. The bulk modulus and mass density of the fluid are $\kappa = 2.2$ GPa and $\rho^F = 1.0 \times 10^3$ kg/m$^3$. **(a)** By fitting the dispersion curve of the sample without pre-stretch, the initial shear moduli of the phantom gel is obtained as $\mu_0 = 37$ kPa. The neo-Hookean model (the only material parameter is $\mu_0$) then provides good predictions for the other dispersion curves with stretches of $\lambda = 1.08, 1.18$. **(b)** A destructive tensile test on the gel gives $\mu_0 = 36$ kPa. **(c)** By fitting the stiffening parameter $b$ of the Fung model on the curve obtained with stretch $\lambda = 1.18$, a very good agreement is found for the prediction of the curve with stretch $\lambda = 1.08$. **(d)** By fitting all three material parameters of the fourth-order elasticity model on the on the $\lambda = 1.18$ curve, a very good agreement is found for the prediction of the curves with stretches $\lambda = 1.00$ and $\lambda = 1.18$.

## 5. Discussion

The waves induced by mechanical and acoustical stimuli in thin-walled biological soft tissues such as mitral valve, cornea, artery and bladder, are *guided waves* (Han et al., 2015). These guided waves are *dispersive* and the mechanical properties of the thin walls cannot be directly inferred from their velocities, in contrast to the situation for bulk shear waves, travelling at a speed independent of the frequency. Instead, a *dispersion analysis* has to be carried out to understand the correlation between the phase velocity and the frequency. Moreover, most soft tissues contain *pre-stresses* in their *in vivo* state and may undergo finite deformations. Thus it is necessary and important to analyze the dispersion equations within the framework of finite deformation theory and to incorporate the effects of pre-stresses. The theoretical solutions derived here serve this purpose. When the level of pre-stress is known, *material parameters* such as the initial shear modulus and hardening parameters can be inferred from the dispersion curves.

The method developed here may find some clinical applications. Indeed, in clinics the accurate determination of local blood pressure in a non-invasive manner is an ongoing pursuit. Variation of the blood pressure in the arteries causes clear deformation of the arterial wall, which can be observed from the ultrasound image (Ribbers et al., 2007). However, the knowledge on the deformation alone is not sufficient for determining the local blood pressure without the prior knowledge of the mechanical properties of the arterial wall. The theoretical solutions presented here provide a promising means to deal with this challenging issue. In particular, once the real-time dispersion relations of guided circumferential waves are measured and the local deformation (i.e., the circumferential stretch ratio) is extracted from the ultrasound image, the present theoretical solutions can be used to estimate the variation of the local blood pressure. Briefly, it is possible to measure the stretch ratio along the circumferential direction using ultrasound imaging (Ribbers et al., 2007). When the circumferential stretch ratio is measured and the dispersion relation of the guided circumferential wave is obtained at the same time, the theoretical model presented in this study can be invoked to determine the elastic modulus of the arterial wall. Then from the known elastic modulus and the circumferential stretch ratio, the blood pressure can be estimated. To realize the measurements in practical cases, we may however have to consider the features of real arteries. Indeed, it has been long recognized that arterial walls are multi-layered structures and that each layer may be considered as a fiber-reinforced composite (Holzapfel et al., 2000). If the artery is modelled as a layered tube, more general incremental theory (Ogden, 2007) should be used to addressed the inhomogeneous prestress within the arterial wall, but the analytical dispersion relation similar to Eq. (25) is really hard to be obtained. When the theoretical solution given by Eq. (25) is used, arterial modulus determined using the guided wave elastography represents the effective modulus of the arterial wall and the averaged stretch ratio along the circumferential direction should be used. Besides, the effective shear modulus should be understood as the specific value of shear modulus along a given direction, e.g., the circumferential direction. Finally, it must be pointed out that real arteries are surrounded by other soft tissues, and not necessarily by a fluid on both sides as here. Such a simplification is reasonable when the surrounding tissues are much softer than the arterial walls (Couade et al., 2010, Li et al., 2017b), but here the critical modular ratio between the arterial wall and the surrounding soft tissue has not been determined, which deserves further investigation.

## 6. Conclusions

Thin-walled soft tissues (including mitral valve, artery, cornea and bladder) and thin-walled artificial soft biomaterials in their working state usually contain pre-stresses and measuring their mechanical properties in a non-invasive and non-destructive manner remains a challenging issue. This study aimed at addressing this challenge and the following key results were obtained.

First, we performed a theoretical analysis based on nonlinear elasticity theory and incremental motion theory to investigate the propagation of guided waves generated by focused acoustic radiation force in pre-stressed, fluid-loaded hyperelastic plates. For flat plates we derived the dispersion relations analytically. An important insight gained from our analytical solution is that the parameter $\zeta$ which is difficult to determine in ultrasound elastography has negligible effects on the $A_0$ mode.

Second, we built a finite element model to simulate the propagation of the acoustic wave in both flat and curved plates (tubes). The FE results show that the theoretical solutions for plane plates are valid and can be applied to tubular or curved plates, at least when the initial ratio of the curvature radius to the wall thickness of the tube is such that $R_0/(2h_0) \geq 2$.

Third, our theoretical and finite element results enabled the development of an ultrasound-based elastography method to infer the elastic and hyperelastic parameters of pre-stressed thin-walled soft biomaterials. The method may be used to characterize *in vivo* the mechanical properties of soft tissues e.g., artery and bladder, and measure *in situ* the elastic properties of artificial thin-walled soft biomaterials in their working state.

Finally, to validate the theoretical solutions and demonstrate the usefulness of the proposed method in practical measurements, we performed experiments on polyvinyl alcohol (PVA) cryogel phantoms using the Verasonics V1 System equipped with a L10-5 transducer. The results show that the method is valid when the pre-strain in the soft biomaterial is up to at least 18%.

**Acknowledgements**

We acknowledge support from the National Natural Science Foundation of China (Grant Nos. 11572179, 11172155, 11432008, and 81561168023) and from the Irish Research Council.